# Complexity of brain tumors

Miguel Martín-Landrove[a,b,†*], Francisco Torres-Hoyos[c,d,†], Antonio Rueda-Toicen[a,†]

[a]*Center for Medical Visualization, National Institute for Bioengineering, Universidad Central de Venezuela, Caracas, Venezuela*

[b]*Centro de Diagnóstico Docente Las Mercedes, Caracas, Venezuela*

[c]*Department of Physics, Universidad de Córdoba, Montería, Colombia*

[d]*Department of Systems Engineering, Universidad Cooperativa de Colombia, Montería, Colombia*

[†]*Physics and Mathematics in Biomedicine Consortium*

**Abstract**

Tumor growth is a complex process characterized by uncontrolled cell proliferation and invasion of neighboring tissues. The understanding of these phenomena is of vital importance to establish appropriate diagnosis and therapeutic strategy and starts with the evaluation of their complexity with suitable descriptors, such as those produced by scaling analysis. In the present work, scaling analysis is used for the extraction of dynamic parameters that characterize tumor growth processes in brain tumors. The emphasis in the analysis is on the assessment of general properties of tumor growth, such as the Family-Vicsek ansatz, which includes a great variety of ballistic growth models. Results indicate in a definitive way that gliomas strictly behave as it is proposed by the ansatz, while benign tumors behave quite differently. As a complementary view, complex visibility networks derived from the tumor interface support these results and its use is introduced as a possible descriptor in the understanding of tumor growth dynamics.

*Scaling analysis; multifractal systems; complex networks*

**1. Introduction**

Tumors exhibit a complex and irregular geometry due to the uneven spatial distribution of their cells. This irregular geometry appears during their growth processes, and it is apparent in a tumor interface with its host, on a tumor vascular network, and even on a tumor's spatial diffusion through time. Fractal geometry provides a notion of dimension that characterizes these complex and irregular objects. In the case of brain tumors, medical imaging technology has been fundamental for the geometrical analysis and quantification of tumor lesions. Magnetic resonance imaging techniques with standard contrast enhancement, dynamic contrast enhancement, and susceptibility weighting give detailed geometrical information with excellent spatial resolution and quality. When applied to the central nervous system, the precise characterization of tumor geometry, in all of its complexity, makes an important contribution to the understanding of brain tumor pathology. This precise geometric characterization leads to new methods for tumor segmentation and tissue classification in enhanced contrast MRI [1], tumor grading [2, 3], and therapy monitoring [3, 4]. Parameters extracted from the complex tumor growth dynamics [6–10, 11, 12] can be used to validate tumor growth models [13-16] for therapy simulation and prognosis [17]. Also, brain tumor complexity and neural brain complexities can be considered to produce models that estimate neurological implications of tumor resection [18] and neurological disorders [19, 20] due to the presence of brain tumors.

---

[*] Corresponding author. e-mail: mglmrtn@gmail.com.

Fractal dimension has been used to characterize morphological irregularities in cancer pathologies and to assess their grade and malignancy [21,22]. It has been used to establish clear geometrical differences between normal, dysplastic, and neoplastic tissues [23]. In the case of brain tumors, fractal dimension, as box-counting or capacity dimension, has been used for tumor segmentation in brain images [1, 3, 24, 25], tumor grading [2, 3], and assessment of the effects of therapy [3, 4]. In these applications, magnetic resonance images with contrast enhancement [3], susceptibility-weighted MRI (which are known as SWI [2, 4]), and histological brain tumor specimens [26-29] have been evaluated. The magnetic resonance imaging modalities used in these studies have high spatial resolution and provide a proper rendering of tumor lesion features. Fractal dimension can also be extracted from the tumor interface after performing an image segmentation process. Several works [11, 12, 30, 31] analyzed tumor interfaces extracted from contrast-enhanced magnetic resonance images and determined the fractal capacity dimension of the tumor interface. Summarizing, the set of fractal dimensions, each one associated to a feature of the tumor lesion, e.g., contrast agent intensity, image texture, vascularity, and tumor interface, supply an adequate description to characterize the transitions from normal to dysplastic to neoplastic tissue [23]. This description is of great help in diagnosis and therapy monitoring. Fractal capacity dimension is in general very easy to calculate by box-counting or sandbox algorithms, which makes it useful for its extended use in clinical applications and computer-aided diagnosis. However, fractal capacity dimension alone does not adequately describe multifractal systems [30, 32-34], so a more general approach to assess the complex behavior of cancer must be addressed using scaling analysis techniques [6-9, 11, 12, 30].

The sections in this article are organized as follows: first, the use of a scaling analysis approach to estimate growth parameters extracted from tumor interface dynamics and their relation to fractal dimensions stressing the pertinence of the Family-Vicsek ansatz [35] as a discriminator between malignant and benign tumors in brain. In second instance, some multifractal analysis techniques are discussed, e.g., a time-like series derived from the tumor interface's rugosity and its associated complex network are analyzed, visibility networks [36, 37] are used to discriminate between malignant and benign tumors [10].

## 2. Methods

### 2.1. Image analysis

#### 2.1.1. Image selection

Images for high grade gliomas were extracted from different collections in The Cancer Imaging Archive [40, 41]; The Cancer Genome Atlas Low Grade Glioma (TCGA-LGG) data collection [42] and the Repository of Molecular Brain Neoplasia Data (REMBRANDT) [43] for astrocytomas and oligodendrogliomas of grades 2 and 3, and The Cancer Genome Atlas Glioblastoma Multiforme [TCGA-GBM] collection [44] for glioblastoma multiforme. For benign brain tumors, local image datasets were used. Among these collections, T1-weighted images, either contrast enhanced or not, were selected and further reviewed, i.e., tumor lesions should be clearly identified as such and separated from anatomical structures, for image processing.

#### 2.1.2. Image processing and tumor interface extraction

The selected images were subjected to the following processing scheme:

Step 1: Selection on the slices that include observable tumor lesion.

Step 2: If two MRI modalities are available, post and pre contrast T1-weighted MRI, the images are then registered using affine transformations, the One plus One Evolutionary method [45] as optimizer and the Mutual Information [46] as a metric, using the Mattes algorithm [47]. In the case of a single MRI modality, i.e., contrast enhanced T1-weighted, this step is skipped.

Step 3: Regions of interest or ROIs are selected slice by slice surrounding the tumor lesion. Digital levels obtained from these regions are classified according to the k-means algorithm [48]. To estimate the number of possible classes, a clustering method based on quantum mechanics proposed by Horn et al. [49] was used. The method assumes that the Parzen estimator [50] of the data points corresponds to the ground state of the Schrödinger equation and a potential energy can be obtained for that state. Applied to digital levels, the Parzen estimator is given by the convolution of the image histogram with the gaussian kernel,

$$\varphi(X) = h \ast g = \sum_i^N h(X_i) e^{-\frac{1}{2\sigma^2}(X-X_i)^2} \quad (1)$$

where the sum extends over all digital levels in the image and $h$ is its histogram. The potential, $V(X)$, is given by,

$$V(X) = \frac{1}{2\sigma^2} \frac{\nabla^2 \varphi(X)}{\varphi(X)} =$$

$$= \frac{1}{2\sigma^4} \left[ \frac{1}{\sigma^2} \frac{\sum_i^N h(X_i)(X-X_i)^2 e^{-\frac{(X-X_i)^2}{2\sigma^2}}}{\sum_i^N h(X_i) e^{-\frac{(X-X_i)^2}{2\sigma^2}}} - 1 \right] \quad (2)$$

The potential $V$ is in fact a better discriminant [49] of the cluster structure than the Parzen estimator, i.e., there are more minima in the potential function than maxima in the Parzen estimator, as a consequence, the selection of a number of classes that correspond to the number of potential minima provides a secure starting point for the application of the k-means algorithm in the classification of the digital leves in the image. After the classification procedure performed by the k-means algorithm is done, all classes that correspond to tumor activity, i.e., contrast enhanced voxels, are selected by inspection with no restriction of the number of classes involved and the image is segmented accordingly. Some examples of the proposed classification scheme are shown in figure 1.

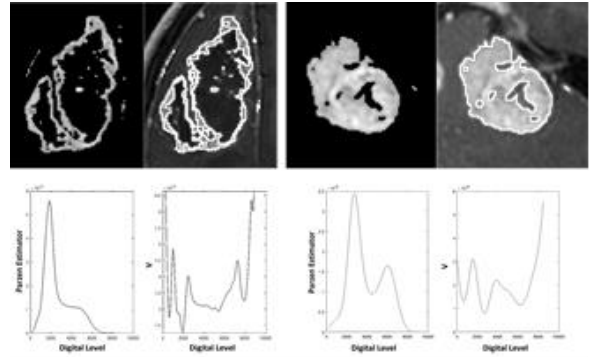

Fig. 1. Examples of results applying the proposed segmentation and classification scheme. On top, segmented images corresponding to glioblastoma multiforme (left) and acoustic schwannoma (right). Bottom, graphics representing the Parzen estimator and its associated Schrödinger potential for both cases.

Step4: Determination of the tumor interface is performed by a method described elsewhere [12] and the results are reviewed slice by slice to exclude structures not related to the tumor interface.

The general procedure is shown schematically in figure 2.

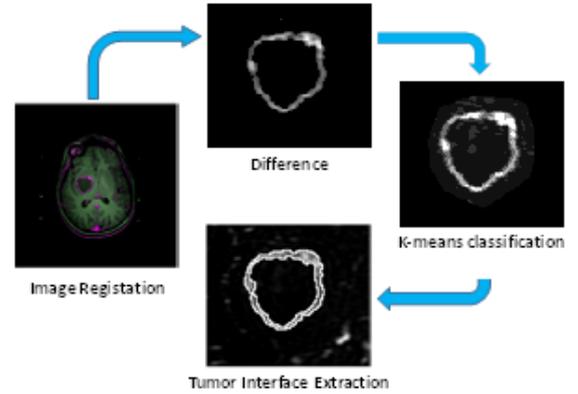

Fig. 2. Schematic representation of the segmentation and classification procedure to obtain the tumor interface. In the case of a single modality acquisition, i.e., contrast enhanced $T_1$-weighted MRI, the first to steps are omitted.

*2.2. The scaling analysis approach*

Besides fractal dimension, there are other ways to describe the fractal geometry of a system. Many other exponents can be derived from the observed power-law behavior through scale transformations. Tumors are complex adaptive systems that can be characterized by dynamics similar to power-law behavior. The growth of tumors, in both resected and in vitro samples, has been characterized using a combination of fractal and scaling analysis techniques [6-9]. These studies have shown that tumor contours exhibit super-rough scaling dynamics described by the Family-Vicsek ansatz [35], which establishes that, for a balistic growth process, the interface width scales as [5],

$$W(L,t) \sim L^\alpha f\left(\frac{t}{L^z}\right) \quad (3)$$

where $L$, is the size of the system and $f(u)$ is a general function, which depends of the particular characteristics of the physical system, which satisfies $f(u) \sim u^\beta$, if $u \ll 1$ and $f(u) \sim c$, a constant, if $u \gg 1$; $\alpha$, the roughness exponent, $\beta$, the growth exponent and $z$, the dynamic exponent are related by $z = \alpha/\beta$. In these studies [6–9] it was demonstrated that this scaling behavior [35] applies at the local as well as the global level. As a consequence, the tumor interface can be also parameterized by a local roughness exponent, $\alpha_{loc}$, besides a global roughness exponent, $\alpha > 1$ [6-9].

In three dimensions, the local roughness exponent relates the scale-averaged width of the interface between tumor and host to the scale of growth given by the area of a spherical cap, $s$, exhibiting a power-law behavior [6-9] for small $s$:

$$W^2(s) \sim s^{\alpha_{loc}} \quad (4)$$

with $W$ given by [9],

$$W(s,t) = \left\{\frac{1}{s}\sum_{r_i \in s}[r_i(t) - \langle r_i \rangle_s]^2\right\}_\Sigma^{\frac{1}{2}} = \{w(s)\}_\Sigma^{\frac{1}{2}} \quad (5)$$

where $\langle r_i \rangle_s$ represents the average radius, measured from the tumor center, over an interface spherical cap of area $s$, and $\{*\}_\Sigma$ represents the average over all realizations (all possible spherical caps of area $s$) over the interface surface $\Sigma$.

In order for the growing process to follow the Family-Vicsek ansatz [35], fractal dimension and local roughness exponent are related in a general way [5, 35], i.e., their sum is equal to the embedding dimension of the shape, or Euclidean dimension, $d_E$,

$$\alpha_{loc} + d_F = d_E \quad (6)$$

Previous studies [38] performed on the tumor interface of contrast-enhanced MRI, using big [39] local and international databases, such as The Cancer Imaging Archive [40, 41] revealed that the condition given by equation (5) holds only for glioblastoma multiforme and high grade gliomas, $\alpha_{loc} = 0.89 \pm 0.08$, $d_F = 2.11 \pm 0.08$ with $\alpha_{loc} + d_F = 3.00 \pm 0.13$ while in the case of meningiomas and other benign tumors, the result of $\alpha_{loc} = 0.76 \pm 0.08$, $d_F = 1.91 \pm 0.06$, with $\alpha_{loc} + d_F = 2.67 \pm 0.11$ is obtained, stressing the fact that both types of tumors exhibit a very different dynamical growth behavior.

Also it is possible to obtain the roughness exponent $\alpha$, which is related to the interface width, $W$, through the power-law behavior [5, 9, 35] and equation (3),

$$W_{sat}(R) \sim R^\alpha \quad (7)$$

where $R$ is the mean radius of the tumor as a measure of its size, and it is assumed that for in vivo tumors, the saturation condition is attained, i.e., that the lateral correlation length for the interface fluctuations is comparable or larger than the size of the system [5, 35]. The results previously obtained [38] for the exponent $\alpha$ were 1.002 for glioblastoma multiforme, 1.262 for metastasis, 0.963 for meningiomas and 0.889 for acoustic schawnnomas. Glioblastomas and metastases exhibit an exponent $\alpha > 1$, which corresponds to a super-rough dynamics for the tumor growth process, denoting the highly invasive character that's typical of malignant neoplastic tissue characterized by high proliferation and diffusion to the tumor interface, in contrast to the case of benign tumors which obey to a growth dynamics of a bulk proliferative process with no diffusion to the tumor interface.

## 2.3. Complex network analysis

Tumor interface growth exhibit multifractal behavior that can not be characterized solely by capacity fractal dimension. More information about the complex dynamics at the tumor interface can be obtained by time series analysis. Brú et al. [51] have established a link between the evolution of complex networks and the dynamical processes that produce rough and fractal-like interfaces. The degree of the nodes in these networks change through time as the interface evolves. The application of this network methodology enables the uncovering of so-called "scale-free" temporal and geometric features that remain invariant as the interface grows. This invariance is detected in the degree distribution of the complex network derived from the interface. This approach could possibly be used to understand tumor interface dynamics.

In particular, visibility graphs defined by Lacasa et al. [52, 53] obtained from the tumor interface points can be used to capture the geometrical correlations that exist among the discrete points that this interface. In general, tumor interface data points are scattered in a two-dimensional array in a very complex manner, as shown in figure 3, which makes streamly difficult the evaluation of the visibility graph in a general way.

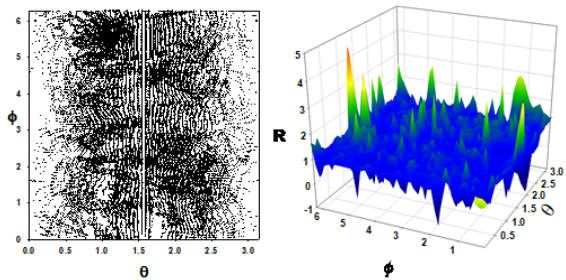

Fig. 3. Example of the complexity of the two-dimensional array of tumor interface data points for a glioblastoma multiforme. On the left, the angular coordinates for each data point and on the right, the complex fluctuations of the radius.

One simplifying assumption is to obtain one-dimensional spatially ordered series derived from geometrical adjacent points over the tumor interface. The simplest approach is to consider spatially ordered series coming from interface points located on the same image slice, which means that these points are ordered according to the angle ɸ, and spatial correlation could be revealed from the ordered series analysis, as shown in figure 4.

The visibility graph is therefore obtained as discussed elsewhere [10, 36, 37] and the associated connectivity or degree distribution is determined for the complete set of slices that comprise the tumor interface. Also, as these ordered series exhibit periodic boundary conditions with respect to the ordering parameter, the connectivity for every single node of the visibility graph can be obtained appropriately.

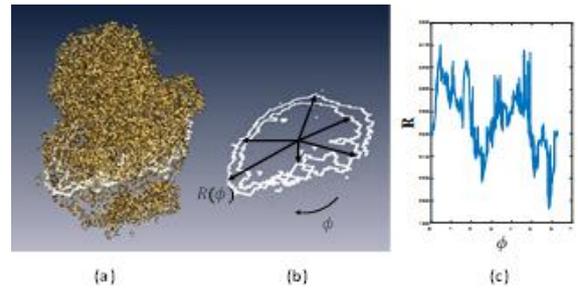

Fig. 4. Extraction of one-dimensional spatially ordered series. (a) Slice localization within the tumor interface, (b) Sampling of interface points $R(\phi)$ and (c) spatially ordered series.

For a network, the connectivity can be established by the simple counting of edges for each node, this count being defined as its degree. The result of this computation for all the nodes in the network is a distribution of degrees. If one considers this distribution as a probability distribution, $P(k)$, it represents how a particular node, selected randomly, is connected to exactly k nodes. The presence of a power law behavior usually denotes a scale-free character for the visibility graph,

$$P(k) \sim k^{\gamma} \qquad (8)$$

It is possibly expected that the exponent $\gamma$ will differ significantly according to the tumor interface irregularity.

## 3. Results and discussion

A total of 295 tumor interfaces were analyzed, discriminated as follows, 130 benign tumors including meningiomas and acoustic schwannomas, 55 Grade II an Grade III astrocytomas and oligodendrogliomas [42,43], and 110 glioblastoma multiforme Grade IV tumors [44]. To extract the tumor interface, all image data sets were segmented by the k-means algorithm using as many classes as predicted by the Schrödinger potential [49], $V$, given by equation (2). In the majority of the cases, the number of classes were between 6 and 8, and among them up to 4 classes were related to tumor activity. Average geometrical properties of the tumor interface are summarized in Table 1.

Table 1. Average geometrical properties of tumor interface. $\langle W \rangle$, average interface width, $\langle R \rangle$, average radius, and $\langle N \rangle$, average number of tumor interface points.

| Tumor group | $\langle W \rangle$ (mm) | $\langle R \rangle$ (mm) | $\langle N \rangle$ |
|---|---|---|---|
| Benign tumors | 2.47 ± 0.84 | 11.92 ± 2.03 | 7691 |
| Grade II and III gliomas | 4.35 ± 1.73 | 15.57 ± 4.34 | 10632 |
| Glioblastoma multiforme | 4.90 ± 1.49 | 19.08 ± 4.14 | 14786 |

Analysis of Table 1 immediately reveals a dramatic change in the average interface width comparing benign tumors to malignant ones, possibly due to an increase of the interface roughness and fractality, associated to proliferative-invasive processes that characterize cancerous tumors.

### 3.1. Scaling analysis results

The results obtained from the scaling analysis are somewhat scattered in the $\boldsymbol{\alpha_{loc} - d_F}$ parameter space as can be seen in figure 5, in which it is evident that the data points are clustered about two well defined classes corresponding mainly to benign and malignant tumors respectively.

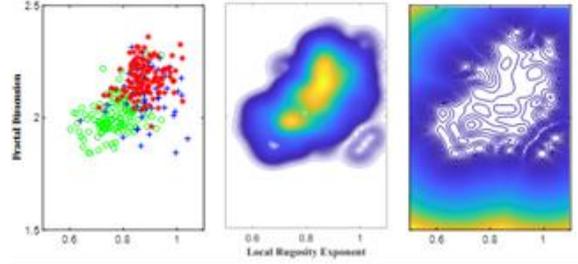

Fig. 5. Left, scatter plot of the results in the $\alpha_{loc} - d_F$ parameter space; green symbols correspond to benign tumors, red to glioblastoma multiforme and blue to grade II and Grade III gliomas; center, Parzen estimator and on the right, the associated Schrödinger potential.

Also, from figure 5, contour plots for the Parzen estimator and its associated Schrödinger potential suggest the possibility of a greater number of classes, i.e., through the number of potential minima [49], corresponding each one to different tumor grades or histologies.

The evaluation of the scaling analysis parameters yielded for glioblastoma multiforme, $\alpha_{loc} = 0.88 \pm 0.06$, $d_F = 2.17 \pm 0.07$, with $\alpha_{loc} + d_F = 3.05 \pm 0.10$, for Grade II and III gliomas, $\alpha_{loc} = 0.89 \pm 0.07$, $d_F = 2.10 \pm 0.11$, and $\alpha_{loc} + d_F = 2.99 \pm 0.13$, and for benign tumors, $\alpha_{loc} = 0.77 \pm 0.07$, $d_F = 2.02 \pm 0.09$, and $\alpha_{loc} + d_F = 2.79 \pm 0.13$, which are in agreement of previous results [38]. The difference in scaling parameters between gliomas with different tumor gradings is not significant, not being this the case when compared to benign tumors. This fact can be supported further if a k-means classification procedure is performed over the $\alpha_{loc} - d_F$ parameter space (see figure 5) assuming only two classes, malignant and benign tumors. The classification yielded for the malignant tumors, i.e., gliomas, independently of grading, $\alpha_{loc} = 0.89 \pm 0.06$, $d_F = 2.18 \pm 0.07$, and $\alpha_{loc} + d_F = 3.07 \pm 0.09$, and for benign tumors, $\alpha_{loc} = 0.77 \pm 0.08$, $d_F = 2.00 \pm 0.06$, and $\alpha_{loc} + d_F = 2.77 \pm 0.10$. Evaluating this classification scheme as a predictive method for diagnosis it has a sensitivity of 0.8364, a specificity of 0.8846, an accuracy of 0.8576 and a precision of 0.9020. The method could be improved noticeable if high resolution images can be afforded by, for example, reducing slice thickness and in plane resolution.

It is important to notice that the variations in the values for the local roughness exponent, $\alpha_{loc}$, and the fractal dimension, $d_F$, determines what proliferative-invasive process describing the dynamics of tumor growth is present. The results obtained by scaling analysis are in agreement to what is to be expected for the geometrical parameters shown in Table 1. As previously reported [38], high grade gliomas and glioblastomas definitely exhibit a balistic growth model, characterized by the ansatz of Family-Vicsek [35], equation (3), while in the case of benign tumors a different grrowth model have to be considered since equation (6) does not apply.

Furthermore, and following equations (3) and (7), the global roughness exponent, $\alpha$, for high grade gliomas and glioblastomas is, respectively, 1.17 and 1.11, which means that the growth process corresponds to a super-rough dynamics compared to a rather smooth growth process for benign tumors with $\alpha = 0.87$. These results are shown in figure 6.

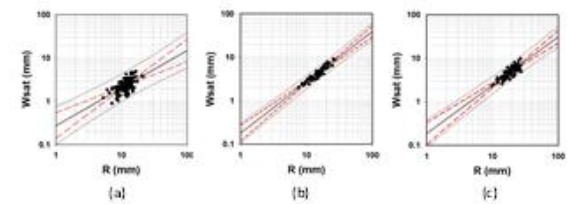

Fig. 6. Dependence of the saturation value of the interface width with tumor size. Lines represent, black continuous, data trend; red segmented, data confidence interval and black dotted, data prediction interval. Exponents are determined assuming equation (7). (a) Benign tumors, α = 0.87, (b) Grade II and III Gliomas, α = 1.17 and (c) glioblastoma multiforme, α = 1.11.

*3.2. Complex network analysis results*

One dimensional spatially ordered series were obtained by the procedure shown in figure 4. Due to restrictions in the size of the series, i.e., some lesions were not big enough for a reliable calculation of the degree distribution, only a subset of the total number of tumor interfaces were considered, 102 benign tumors, 50 high grade gliomas and 101 glioblastomas.

Some examples of the calculated degree distribution, for a meningioma and a Grade III astrocytoma, are shown in figure 7. As can be seen in the figure, individual degree distributions are subjected to fluctuations that depend on the number of points at the tumor interface, so an individual evaluation of the power law exponents depends on the size of the tumor lesion and differences between tumor groups are somewhat diminished. This fact limits the evaluation of the exponent only to large tumor lesions, such as glioblastomas and some high grade gliomas, excluding the majority of benign tumors, according to table 1.

In order to establish a characteristic degree distribution for a tumor group, all the points belonging to the tumor interfaces within the group are considered as nodes of a visibility graph that will be representative of the whole group, under the assumption that tumor interface fluctuations are similar for tumors belonging to the same group.

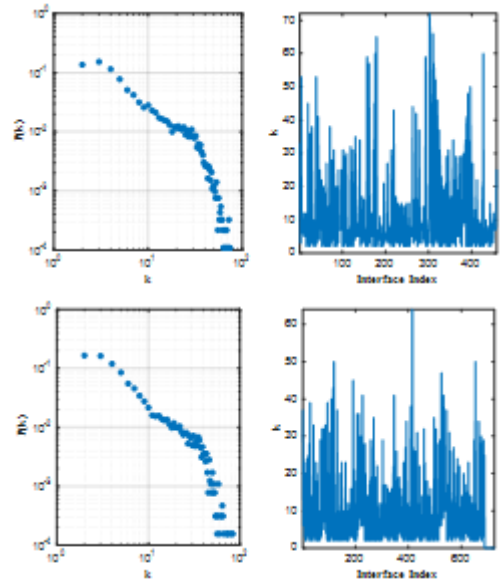

Fig. 7. Examples of the degree distribution $P(k)$ and $k$ series for a meningioma (top) and a Grade III astrocytoma (bottom).

Assuming only three groups, i.e., benign tumors, Grade II and III gliomas and glioblastomas, the group degree distributions are shown in figure 8, all of them exhibit two regions for which a power law behavior can be extracted, the first one in the range of $3 < k \leq 10$ and the second one for $10 < k \leq 20$, just before the onset of the cut off in the degre distribution due to the finite size of the spatially ordered series.

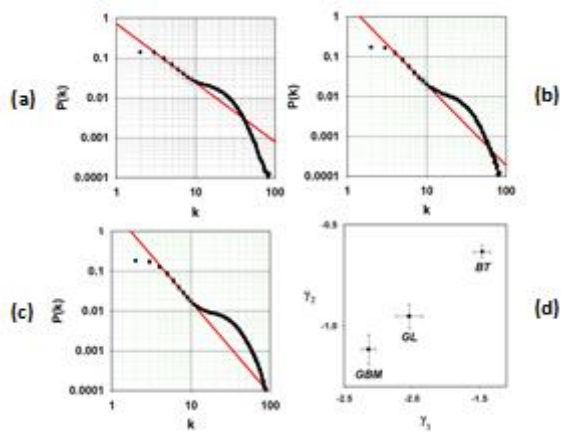

*Fig. 8.* Group degree distributions for (a) Benign tumors, (b) Grade II and Grade III gliomas and (c) glioblastoma multiforme. Red lines indicate the power law fit for the first region. (d) Exponents for the power law behaviors for each tumor type.

Results reveal significant differences between the tumor groups in the exponent for the first power law region, $\gamma_1$, obtaining $-1.48 \pm 0.06$, for benign tumors, $-2.02 \pm 0.10$, for Grade II and III gliomas and $-2.32 \pm 0.05$, for glioblastomas. For the second region the obtained exponents, $\gamma_2$, are $-0.63 \pm 0.03$, for benign tumors, $-0.95 \pm 0.06$, for Grade II and III gliomas and $-1.12 \pm 0.07$, for glioblastomas. Comparing the results obtained for $\gamma_1$ with those published by Brú et al. [10], where different growth models that obey the Family-Vicsek ansatz are analyzed by visibility graphs, there is some correspondence of the Edwards-Wilkinson [51] and Kardar-Parisi-Zhang [52] models, which exhibit $\gamma$ in the range $-2.10$ to $-2.07$ [10], with the result obtained for Grade II and III gliomas, and of the Random Deposition with Surface Relaxation [53] and Eden [54] models, with $\gamma$ in the range $-2.46$ to $-2.25$, with the result for glioblastoma multiforme. In the case of benign tumors the comparison is not possible since, as it was previously disscused, its growth dynamics does not correspond to a balistic deposition growth model and therefore, it is beyond the analysis of reference [10], but nevertheless, this fact adds up to assess a different growth model for this group. The meaning of the second exponent, $\gamma_2$, is still unclear since it is affected by the cut off in the degree distribution due to the size of the spatially ordered series, which is rather small, i.e., the maximum size for the tumor interfaces analized in this work is of the order of 2000. Nevertheless, exponents for high grade gliomas and glioblastomas are more negative than those for benign tumors, a result that is consistent with results for the first exponent $\gamma_1$, and those coming from scaling analysis and Table 1. One possible interpretation is that in malignant tumors, points are dispersed on a wide hiper rough interface, and it is expected that the majority of nodes in the visibility graph are located in deep valleys and as a consecuence the degree distribution falls off more rapidly with k. In the case of benign tumors since they exhibit a smooth interface and its width is smaller than for malignant ones, as shown in Table 1, the number of connections for each node will not differ in much depending on its location on the interface and therefore its degree distribution will fall off more slowly.

It has to be remarked that the spatially ordered series considered in this work and in reference [10] are considered "flat", in the sense that the horizon is always visible no matter the length of the series. Under actual conditions, the tumor interface is a closed surface and therefore points at the interface are hidden by the bulk volume of the tumor. As a consequence, the degree distributions, P(k), are limited to a certain number of neighbours what implies an smaller cut off connectivity for the visibility graph, and possibly characteristic exponents may vary, so the previous correspondence with growth models should be taken within the context of reference [10].

Visibility graphs and corresponding degree distributions were calculated for the different tumor types assuming a closed tumor interface, and the degree distributions are shown in figure 9, comparing benign tumors and glioblastoma multiforme.

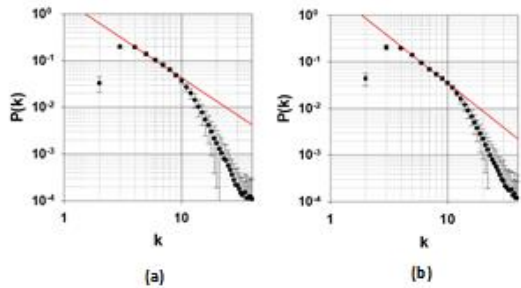

*Fig. 9.* Degree distributions for visibility graphs considering "curved" spatially ordered series. (a) Benign tumors and (b) glioblastoma multiforme.

Since a fast fall off in the connectivity is expected only one exponent $\gamma$ was considered. The obtained exponents were $-1.67$ for benign tumors and $-1.99$ for glioblastoma multiforme, showing the same tendency as previous results.

## 4. Conclusions

It is presented a general methodology to extract tumor interfaces from contrast enhanced MRI by a classification scheme that includes the assessment of the number of classes given by the number of minima of a quantum mechanically based Schrödinger potential for its use in a k-means classification algorithm. The extracted interfaces, belonging to different tumor groups including meningiomas, acoustic schwannomas, Grade II and III astrocytomas and oligodendrogliomas, and glioblastomas, were evaluated in relation to its geometrical properties and dynamics. It was demonstrated that for gliomas, independently ot its grade, the tumor growth dynamics completely adjust to a growth process that obey the Family-Vicsek ansatz, while for benign tumors, a different growth model has to be proposed. The use of complex visibility networks add some support to this result in a consistent manner, but limited by the tumor interface size. This will require some improvements in image acquisition in relation to spatial resolution. Further work is necessary to exploit the full capacity of complex networks analysis in extracting the dynamic characteristics of tumor growth, a task that will be acomplish in the near future.

**Acknowledgments**

The authors would like to thank the financial support and resources received from their respective affiliations in the realization of this work and also the collaborative integration provided by the Physics & Mathematics in Biomedicine Consortium. The results shown here are in whole or part based upon data generated by the TCGA Research Network: http://cancergenome.nih.gov/.